\documentclass[fleqn,usenatbib]{mnras}

\usepackage{newtxtext,newtxmath}

\usepackage[T1]{fontenc}

\DeclareRobustCommand{\VAN}[3]{#2}
\let\VANthebibliography\thebibliography
\def\thebibliography{\DeclareRobustCommand{\VAN}[3]{##3}\VANthebibliography}

\usepackage{graphicx}	
\usepackage{amsmath}	

\title[15 Vul, an Evolved $A_{m}$ Star]{A Critical Evaluation of the Elemental Abundances and Evolutionary Status of 15 Vul (HD 189849): The First Identification of an Evolved $A_{m}$ Star}

\author[M. T. \c{C}ay et al.]{
M. Ta\c{s}k{\i}n \c{C}ay $^{1}$\thanks{E-mail: taskin@istanbul.edu.tr}
\.{I}pek H. \c{C}ay $^{1}$
and
Bet\"{u}l Civelekler $^{2}$
\\
$^{1}$ \.{I}stanbul University, Faculty of Sciences, Astronomy and Space
Sciences Department, 34119, Beyaz{\i}t, \.{I}stanbul, Turkey\\
$^{2}$ Unemployed astronomer, not affiliated with an institution.\\
}

\date{Accepted XXX. Received YYY; in original form ZZZ}

\pubyear{\the\year{}}

\begin{document}
\label{firstpage}
\pagerange{\pageref{firstpage}--\pageref{lastpage}}
\maketitle

\begin{abstract}
Understanding the evolution of metallic-line ($A_{m}$) stars requires well-determined atmospheric parameters and abundance patterns of the selected candidates. In this study, we presented a detailed abundance analysis of 15 Vul (HD 189849), identified as a marginal $A_{m}$ star, using a combination of equivalent width and spectrum synthesis techniques, under the LTE assumption. We compared our findings to previous analyses of the star, providing critique on both their results and our own. Our results suggest that although 15 Vul exhibits some underabundances of calcium and scandium, which are typically associated with $A_{m}$ stars, it might be more accurately identified as a normal A star in terms of its abundance pattern of all other elements and microturbulence velocity. The star’s position on the HR diagram, along with our findings, may indicate that it is potentially a classical $A_{m}$ star that has evolved into the subgiant phase as a `quasi-normal' star. This may be the first identification of an evolved $A_{m}$ star.
\end{abstract}




\section{Introduction}

Metallic-line stars ($A_{m}$ stars), first identified by \citet{tit40}, represent a distinct class of A to early F-type Population I stars. Their spectra are characterized by stronger absorption lines of heavy elements and a deficiency in calcium and scandium compared to normal A stars with similar hydrogen line strengths. These stars present a notable difference in the inferred spectral types based on the \ion{Ca}{ii} K-lines, hydrogen lines, and/or metal lines. Stars that exhibit a difference in classification between the Ca II K-lines and metal lines of less than five subclasses were originally termed `Proto $A_{m}$' stars by \citet{mor78} and are now known as `marginal $A_{m}$' stars.

The abundance anomalies of $A_{m}$ stars are best explained by diffusion theory \citep{mic70,wat70,ric00}, although further improvements are needed for a complete explanation. These anomalies are caused by the separation of elements through the competition between radiative acceleration and gravitational settling. While this separation also occurs in normal A stars, meridional circulation driven by high rotational velocity homogenizes their atmospheres \citep{cha93}. In contrast, $A_{m}$ stars, which are all slow rotators \citep{abt71}, lack such a mixing mechanism, leading to the observed overabundances and underabundances. Slow rotation causes the helium convection zone to sink and peculiar abundances, driven by radiative separation at the bottom of the hydrogen convection zone, are mixed into the atmosphere through an overshooting mechanism. However, because calcium and scandium have noble gas configurations in this zone, they are not pushed upward and, as a result, are deficient in the photospheres of $A_{m}$ stars.

Since $A_{m}$ stars are generally young objects, it is an intriguing question what happens to them after the main sequence. It has been proposed that $\rho$ Pup stars might be the descendants of $A_{m}$ stars, but the origin of abundance anomalies in these stars is still under debate \citep{gah20}. A challenge to this hypothesis comes from the statistically low occurence frequency of these stars, which would be higher if they were indeed evolved $A_{m}$ stars \citep{abt17} \citep[but see][]{gah20}. In this context, it is noteworthy that \citet{leb08} predicted that, regardless of the details of the scenario in which the depletion of calcium and scandium arises, this depletion will persist into the subgiant phase, while the abundances of other elements will depend on the extent of the mixing zone's depth.

To understand the evolution of $A_{m}$ stars, it is crucial to obtain carefully determined atmospheric parameters and abundance patterns, especially from critically selected targets. In this context, we chose to analyze 15 Vul (HD 189849), as it may be one of the most suitable candidates. It has been reported as a marginal $A_{m}$ star, exhibiting very mild overabundances and also appears to be slightly evolved \citep{far71,ade97}.

\section{The Observation and The Data Reduction}

The observation of the star took place on 2012 August 1, during a single observing session, utilizing the Coud\'{e} \'{e}chelle spectrograph attached to the 1.5-meter Russian-Turkish Telescope (RTT-150) at the T\"{U}B\.{I}TAK National Observatory (TUG) in T\"{u}rkiye. The dataset comprised 5 high-resolution spectra taken with a resolving power (R) of 40000 and a high signal-to-noise ratio (S/N $\geq$ 100), covering the wavelength range of 3800 to 10000 {\AA}. The detector was thermo-electrically cooled (-60 \degr C) {\sc andor ccd (dw436-bv)}, with 2048$\times$2048 pixel chip and 13.5 micron pixel size. Standard preliminary reduction procedures were implemented, including bias and dark subtractions, flat fielding, scattered light correction, and extraction of 1D spectra from the \'{e}chelle orders. The spectra were co-added to improve the S/N ratio. All procedures were performed using {\sc m}ax{\sc i}m {\sc dl} \footnote{\url{https://diffractionlimited.com/product/maxim-dl/}} and {\sc dech} \footnote{\url{https://www.sao.ru/hq/coude/galazut.htm}} \citep{gal22} software packages. {\sc iraf}'s \citep{tod86} spectool package was used for both continuum fitting and equivalent width measurements.

\section{The Analysis}

\subsection{Determination of the Atmospheric Parameters}
\label{sec:parameters}

Due to the Balmer lines' wings extending beyond the limits of \'{e}chelle orders in our spectra, we opted not to utilize these lines for estimating $\textit{T}$\textsubscript{{\sc eff}}  and/or log {\textit g}. Instead, the star's atmospheric parameters were directly derived from its spectrum, establishing the excitation and ionization balances of iron. This can be achieved through the standard procedure of abundance determination by equivalent width analysis. The process involves making a preliminary estimation of atmospheric parameters to set an initial model atmosphere. Subsequently, the model parameters are iteratively adjusted until both the excitation and ionization balances are achieved. To get the preliminary atmospheric parameters we chose to rely on Str\"{o}mgren color indices and \textit{uvby$\beta$} code from \citet{nap93}, with photometric colors obtained from \citet{hau98} through the VizieR facility of CDS \footnote{\url{https://cds.u-strasbg.fr}\label{cds}} and estimated the $\textit{T}$\textsubscript{{\sc eff}} as 7870 K and the log {\textit g} as 3.62. Additionally, IRFM calibrations for Str\"{o}mgren \(b-y\) and Johnson \(B-V\) colors \citep{mel03} yielded effective temperatures of $\textit{T}$\textsubscript{{\sc eff}} = 7690 K and $\textit{T}$\textsubscript{{\sc eff}} = 7650 K, respectively.

\begin{figure}
	\includegraphics[width=\columnwidth]{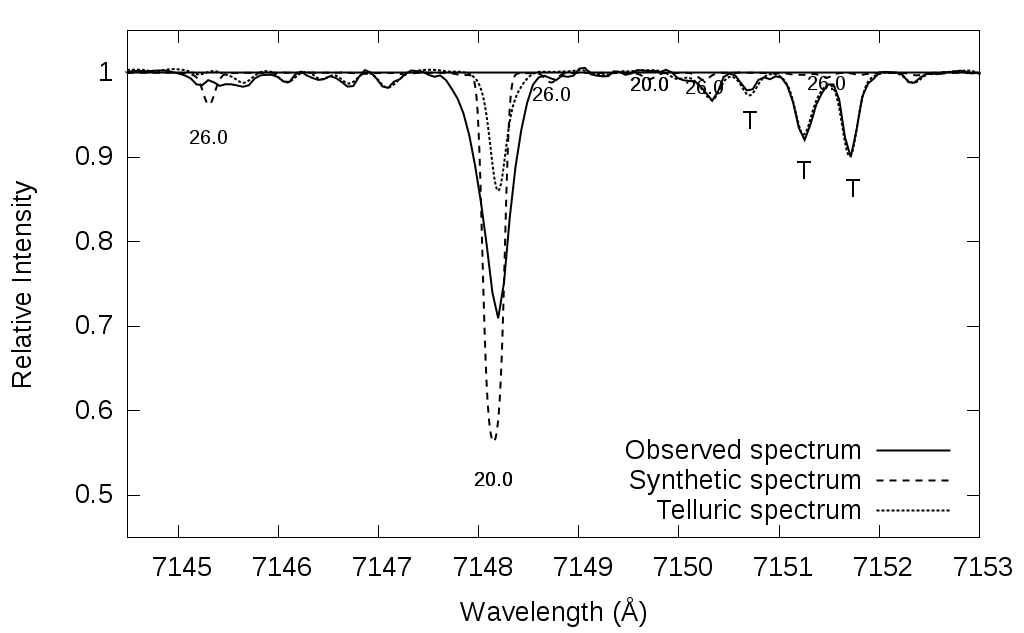}
    \caption{The figure illustrates an instance of the line identification and selection process. The unbroadened synthetic spectrum (dashed line) is utilized for identification purposes. The telluric spectrum (dotted line) matches the resolution of the observed spectrum (continuous line). The elements are identified with the atomic numbers in the figure. In this particular case, the \ion{Ca}{i} line at 7148 {\AA} is contamined by a telluric line. Some additional telluric lines are indicated with letter `T'.}
    \label{fig:fig01}
\end{figure}

The selection of iron lines, along with other elemental lines for analysis, involved a visual process where unblended \ion {Fe}{i} and \ion {Fe}{ii} lines were chosen by overlaying an unbroadened synthetic spectrum onto the observed spectrum. This was achieved by first calculating an LTE model atmosphere using Kurucz's {\sc {\sc atlas9}} code \citep{kur93,sbo07} with specified initial $\textit{T}$\textsubscript{{\sc eff}} and log {\textit g} values. Subsequently, an unbroadened synthetic spectrum was generated for these parameters using the {\sc spectrum} suite \citep {gra94}. Line identification  was carried out by  extracting lists of lines from {\sc spectrum}'s native list and the VALD3 \footnote{\url{http://vald.astro.uu.se}\label{VALD3}} database \citep{pis95,rya97,kup99,kup00,rya15,pak19}, with consideration given to the relevant $\textit{T}$\textsubscript{{\sc eff}} and log {\textit g} values. These lists were then mapped onto both the observed and synthetic spectra. Additionally, a telluric line spectrum extracted from the Solar Flux Atlas of \citet{wal11} was also overlaid after reducing its resolution match to that of our spectra. This process enabled both the estimation and assessment of line blending with nearby lines and potential contamination by telluric lines. Fig.~\ref{fig:fig01} provides an example illustrating the line identification and selection process.

Equivalent widths were measured using  {\sc iraf}’s spectool package, and when required, the deblending option was applied for slightly blended cases. The gaussian fitting technique was employed for equivalent width measurements. Lines with equivalent widths larger than 100 m{\AA} and smaller than 5 m{\AA} were excluded from the analysis because strong lines are sensitive to microturbulence velocity, and weak lines are more susceptible to errors due to inaccurate positioning of the continuum. Abundances from the equivalent widths were calculated using the {\sc abundance} routine of the {\sc spectrum} suite.

\begin{figure}
	\includegraphics[width=\columnwidth]{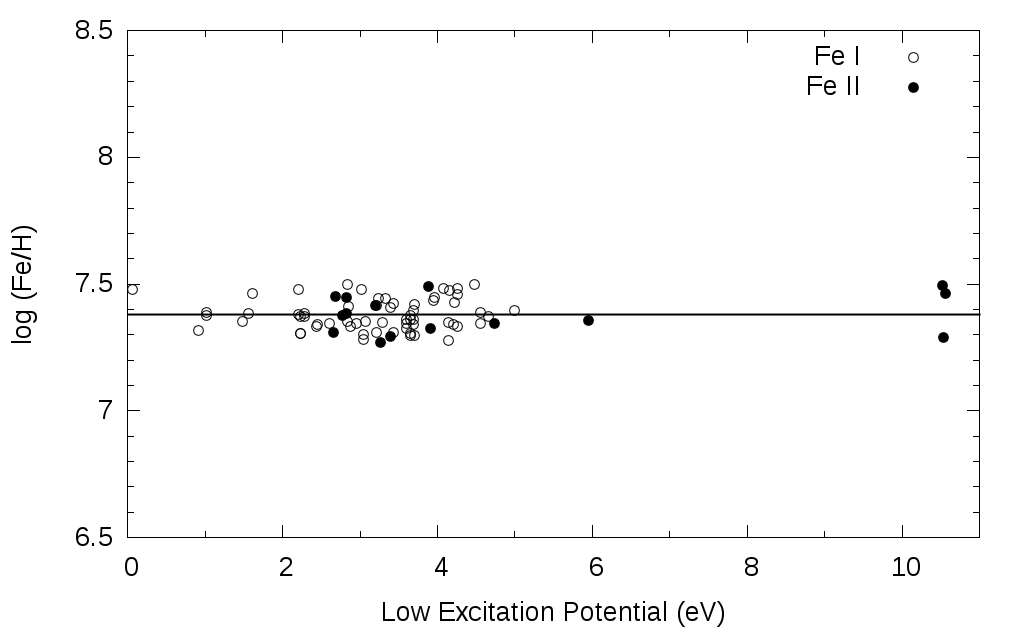}
    \caption{Determination of the iron abundance and fixing the final atmospheric parameters. Figure shows the ionization balance for \ion{Fe}{i} (open circles) and \ion{Fe}{ii} (filled circles). See the text for details.}
    \label{fig:fig02}
\end{figure}

Fig.~\ref{fig:fig02} displays the outcomes of determining the abundance of iron using the equivalent width method. Starting with the initial atmospheric parameters, we carried out iterative abundance calculations using a series of model atmospheres with the goal of achieving two simultaneous conditions: first, the abundances of \ion {Fe}{i} lines became independent of excitation potentials with minimal scatter around the mean; second, the abundances became independent of reduced equivalent widths with minimal scatter around the mean. The former condition determined the effective temperature, while the latter provided the microturbulence velocity. Concurrently, adjusting the log {\textit g} value of the model until the iron abundance from \ion {Fe}{ii} lines equaled that from \ion {Fe}{i} lines established the surface gravity of the model. Consequently, the final model is characterized by $\textit{T}$\textsubscript{{\sc eff}} = 7685 $\pm 100$ K, log {\textit g} = 3.09 $\pm 0.2$  and $\xi$ = 1.95 $\pm 0.5$ km s$^{-1}$. 

\begin{figure}
	\includegraphics[width=\columnwidth]{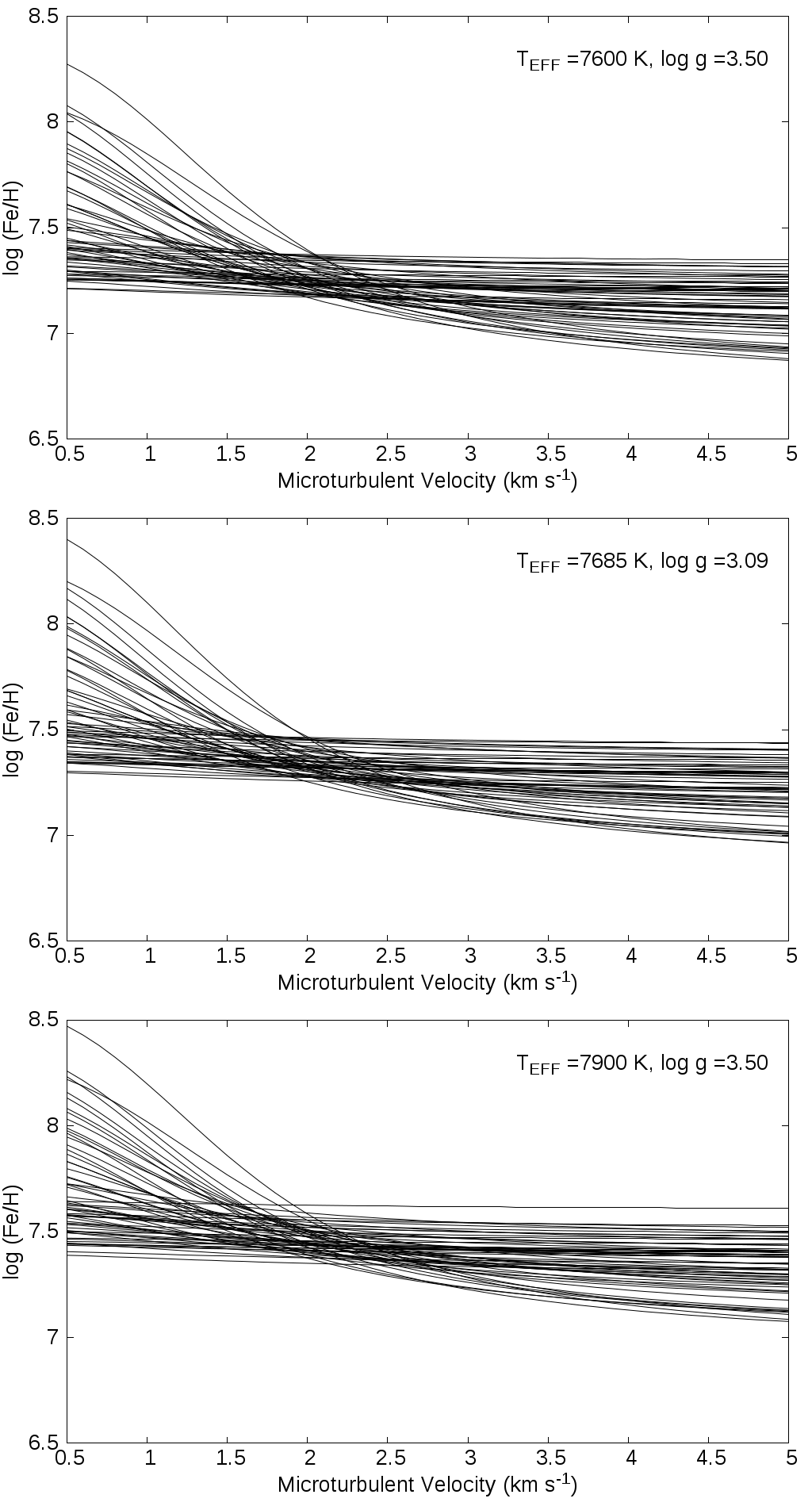}
    \caption{A selection of Blackwell diagrams encompassing our highest and lowest initial $\textit{T}$\textsubscript{{\sc eff}} estimates with log {\textit g} = 3.50 (upper and lower panels) as well as the values of our final model (middle panel). All consistently gives $\xi$ $\approx$ 2 km s$^{-1}$.}
    \label{fig:fig03}
\end{figure}

Because the microturbulence value of the final model is lower than those previously reported for the star and generally accepted for $A_{m}$ stars, we independently verified the microturbulent velocity creating Blackwell diagrams \citep{bla79} for the \ion{Fe}{i} lines. To accomplish this, we generated diagrams using {\sc spectrum}’s {\sc blackwell} routine across a range of models. These models encompassed upper and lower values based on our initial estimates of effective temperature, with log {\textit g} set to 3.50 as the literature value and 3.09 as our finding. Fig.~\ref{fig:fig03} illustrates the results, confirming that the microturbulent velocity is approximately 2 km s$^{-1}$. 

Due to the negligible NLTE departure coefficients for iron lines at the model's temperature and surface gravity (T. Sitnova 2019, private communication), one can anticipate that LTE analysis should yield a reliable abundance result for iron, in turn ensuring internally consistent atmospheric parameters.

\subsection{Testing of the Atomic Data and Determination of the Elemantal Abundances}
\label{sec:atmospheric}

The determination of elemental abundances involved the calculation for 29 different elements, 5 of which exhibit measurable lines in both neutral and singly ionized states. In the analysis, we applied both the equivalent width technique and the spectrum synthesis technique. The spectrum synthesis technique was preferred when a species possesses only one or two distinct measurable lines, and statistical analysis is deemed unsuitable. This preference extends to scenarios where the lines demonstrate hyperfine structure and/or isotopic shifting.

Atomic line parameters were carefully chosen from the literature, and the VALD3 and NIST \footnote{\url{https://www.nist.gov/pml/atomic-spectra-database}\label{NIST}} databases were extensively utilized for this purpose. Experimental oscillator strengths were prioritized whenever they were accessible for calculations. When employing the spectrum synthesis technique, if a spectral line is identified in both the target star and the Sun, the chosen log {\textit {gf}} value was initially tested on the solar spectrum in a way that the calculated abundance from the solar line matches the solar abundance of the relevant element, as provided by \citet{gre98}. We chose to compare our results with the solar abundances from \citeauthor{gre98} because more recent studies use 3D hydrodynamic models in their calculations. The results generated by these models may not be suitable for direct comparison to our 1D LTE calculations. Only in a few instances have minor adjustments been applied to the oscillator strength of a line, either through the inverse analysis technique or simply by hand. This was done with the aim of achieving an improved synthetic line that more accurately represents the corresponding solar line.

The same testing approach was applied to hyperfine structure (HFS) data and van der Waals damping parameters. Hyperfine splitting of a line due to nucleon-electron interaction is a crucial consideration in stellar abundance studies, particularly for odd-Z elements. Ignoring this effect may lead to miscalculations of the abundance of the relevant element  \citep[see, e.g.][]{wah05,jof17}. HFS data was collected from the literature. van der Waals damping parameters, sourced from the VALD3 database, were specifically applied to lines with strong wings. The pertinent literature data is presented in the Table~\ref{tab:tab01}.

\begin{table}
	\centering
	\caption{The lines and the related data used in the analysis with individual abundance results. Explanation of the columns are follows: $\bf{Ion}$, neutral or ionic state of the element used for the analysis; $\bf{Wave.}$, wavelength of the line; $\bf{LEP}$, low excitation potential; $\bf{log {\textit {gf}}}$, logarithm of weighted oscilator strength; $\bf{Mtd.}$, Analysis method ($\bf{E}$quivalent $\bf{W}$idth or $\bf{S}$pectrum $\bf{S}$ynthesis); $\bf{EW}$, equivalent width; $\bf{\epsilon}$ abundance relative to sun ($\bf{\epsilon}$ = $\log({N_{elm.}/H})_*$ - $\log({N_{elm.}/H)_{\sun}}$); $\bf{Ref.}$, references to the log {\textit {gf}} data, The letter references written as superscript are for the HFS data. Full table is available online.}
	\label{tab:tab01}
	\begin{tabular}{llllllll} 
	\hline
\bf{Ion}	& \bf{Wave.} &	\bf{LEP}	&	\bf{log {\textit {gf}}}	& \bf{Mtd.}	&	\bf{EW} & \bf{$\epsilon$}	&	\bf{Ref.}	\\
    & \bf{(\AA)} & \bf{(eV)} &    &   & \bf{(m{\AA})}  &  & \\
\hline
Li I  &  $6707.760^a$  &  0  &  -0.002  &  SS  &  -  &  2.10  &  1\\
Li I  &  $6707.910^a$  &  0  &  -0.303  &  SS  &  -  &  2.10  &  1 \\
C I  &  4770.027  &  7.483  &  -2.437  &  EW  &  33.54  &  0.02  &  2 \\
C I  &  4775.898  &  7.488  &  -2.163  &  EW  &  45.41  &  -0.07  &  2 \\
C I  &  4932.049  &  7.685  &  -1.658  &  EW  &  67.07  &  -0.17  &  2 \\
C I  &  5023.843  &  7.946  &  -2.210  &  EW  &  16.53  &  -0.24  &  2 \\
C I  &  5380.337  &  7.685  &  -1.616  &  EW  &  73.14  &  -0.13  &  2 \\
C I  &  6010.680  &  8.640  &  -1.938  &  EW  &  14.30  &  -0.08  &  2 \\ 
C I  &  6014.840  &  8.643  &  -1.584  &  EW  &  20.59  &  -0.25  &  2 \\
...&...&...&...&...&...&...&...\\
\hline	
\end{tabular}
\bf {References to log \textit {gf} data:}
1: \citet{yan95}
2: \citet{kra21}
3: \citet{ral10}
4: \citet{kur95}
5: \citet{kur07}
6: \citet{kra23}
7: \citet{bie93}
8: \citet{kur04}
9: \citet{smi75}
10: \citet{smi81}
11: \citet{sea94}
12: \citet{law19}
13: \citet{law13}
14: \citet{woo13}
15: \citet{biz93}
16: \citet{woo14}
17: \citet{kur10}
18: \citet{sob07}
19: \citet{mar88}
20: \citet{law17}
21: \citet{den11}
22: \citet{fuh06}
23: \citet{kur08}
24: \citet{wla14}
25: \citet{fuh88}
26: \citet{kur03}
27: \citet{koc68}
28: \citet{war68}
29: \citet{lam69}
30: \citet{han82}
31: \citet{pit86}
32: \citet{lju06}
33: \citet{law01}
34: \citet{law09}
35: \citet{den03}
36: \citet{meg75}
37: \citet{law06}
38: \citet{lwd01}
G: Guess/inverse analysis

\bf{References to HFS data: }
a: \citet{kur95} 
b: \citet{woo14}
c: \citet{arm11}
d: \citet{den11}
e: \citet{pro00}
f: \citet{din91}
\end{table}

For all these tests, the high resolution (R = 384 $\times$ 10 $^{3}$ to 698 $\times$ 10$^{3}$) Solar Flux Atlas of \citet{wal11} served as the observed solar spectrum. The solar model, calculated using {\sc atlas9} with atmospheric parameters set at $\textit{T}$\textsubscript{{\sc eff}} = 5777 K, log {\textit g} = 4.44 and $\xi$ = 0.9 km s$^{-1}$, was obtained from Fiorella Castelli’s website \footnote{\url{https://wwwuser.oats.inaf.it/castelli/sun.html}}.

\begin{figure}
	\includegraphics[width=\columnwidth]{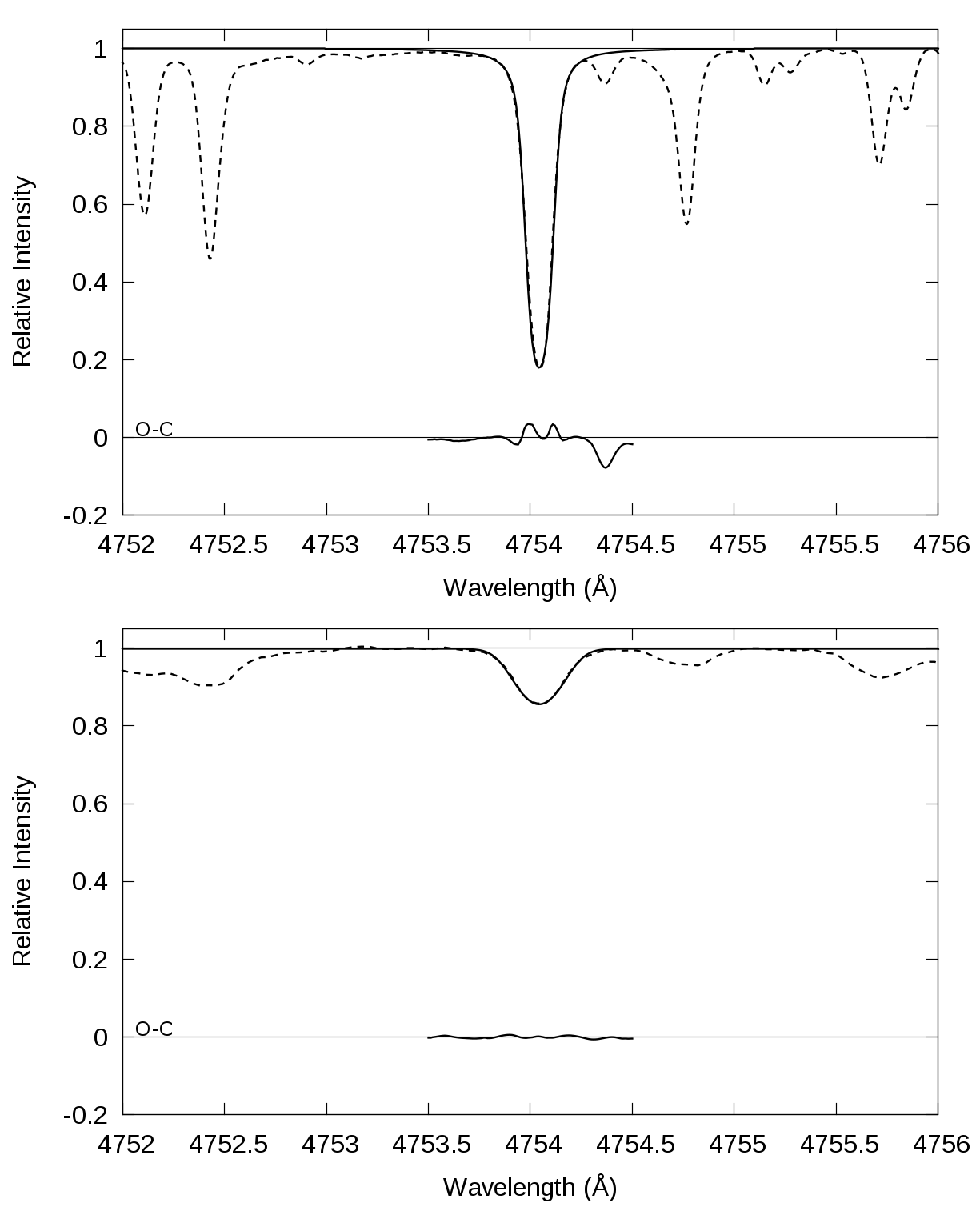}
    \caption{{\bf{Upper Panel}:} Testing of the relevant atomic line parameters on the solar  \ion{Mn}{i} 4754 {\AA} line. The dashed line represents the observed solar spectrum, while the continuous line depicts the synthetic line. The difference between the observed and calculated lines (O-C) is shown. Slight fluctuations of the O-C on both sides of the line near the core maybe attributed to the asymmetry caused by turbulent motions in the solar atmosphere, which are clearly visible in a very high-resolution spectrum like this. Despite these fluctuations, the match between the observed and calculated lines is nearly perfect. {\bf{Lower Panel:}} Synthesis of the stellar  \ion{Mn}{i} 4754 {\AA} line using the accepted atomic line parameters validated on the Solar spectrum. The dashed line represents the observed stellar spectrum, while the continuous line shows the synthetic line. The O-C plot demonstrates an excellent match between the observed and synthetic lines.}
    \label{fig:fig04}
\end{figure}

Fig.~\ref{fig:fig04} illustrates the synthesis of the \ion{Mn}{i} 4754 {\AA} line in the solar spectrum for atomic data testing and in the spectrum of 15 Vul for abundance determination, providing an example.

The agreement between the observed and calculated profiles was determined using one or a combination of methods, including chi-square minimization, point-by-point O-C plot analysis, and comparison of the areas under the observed and calculated lines, depending on the specific circumstances.

As the equivalent width technique was applied to species with multiple measured lines, no testing was conducted on the atomic data. Instead, reliance was placed on the statistical confidence of the results, as is customary. The technique is straightforward, as explained in section~\ref{sec:parameters}.

\subsection{Results of the Abundance Analysis}

\begin{figure}
	\includegraphics[width=\columnwidth]{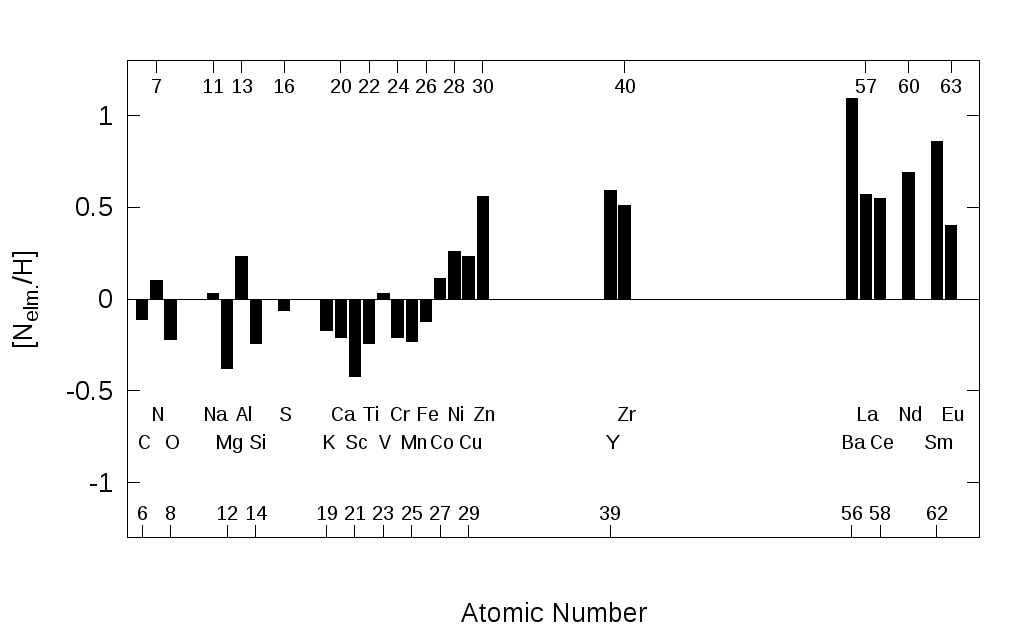}
     \caption{Summary of the mean abundances relative to the Sun.}
    \label{fig:fig05}
\end{figure}

The results of the analysis are presented in Table~\ref{tab:tab02a} and summarized in the Fig.~\ref{fig:fig05}. If the abundance of an element differs between its neutral and singly ionized stages, the accepted mean abundance of the element is calculated giving weight to measured line numbers as follows:

\begin{equation}
    \epsilon_*=\frac{\sum{n_{i}\epsilon_i}}{\sum{n_{i}}},
	\label{eq:quadratic}
\end{equation}
where $\epsilon_i$ represents the abundance from ${i}^{th}$ state, and $n_{i}$ denotes the measured line numbers for the respective state.

\begin{table}
	\centering
	\caption{Abundance analysis results for 15 Vul. Explanation of the columns are follows: $\bf{Elm.}$, symbol of the element; $\bf{Ion}$, neutral or ionic state of the element used for the analysis; $\bf{n}$, number of lines of the element used in calculation; $\bf{\epsilon}$, abundance from relevant state of the element ($\bf{\epsilon}$ = $\log({N_{elm.}/H})_*$ - $\log({N_{elm.}/H})_{\sun}$); $\bf{\sigma}$, standard deviation; $\bf{Mtd.}$, Analysis method ($\bf{E}$quivalent $\bf{W}$idth or $\bf{S}$pectrum $\bf{S}$ynthesis); $\bf{\epsilon_*}$, accepted elemental abundance from the current analysis ($\bf{\epsilon_*}$ same as $\bf{\epsilon}$); $\bf{\epsilon_{\sun}}$, solar abundances from \citet{gre98} ($\bf{\epsilon_{\sun}}$ = $\log({N_{elm.}/H}$)+12).}
	\label{tab:tab02a}
	\begin{tabular}{lccccccc} 
	\hline
\bf{Elm.}	& \bf{Ion}	&	\bf{n}	&	\bf{$\epsilon$}	& \bf{$\sigma$}	&	\bf{Mtd.}	&	\bf{$\epsilon_*$}	&	\bf{$\epsilon_{\sun}$}	\\
\hline															
Li  &       &       &       &       &   SS &    2.10    &   1.10    \\
    &   \ion{Li}{i} &   1   &   2.10    &               &       &       \\
C	&		&		&		&		&	EW	&  -0.11	&   8.52	\\
	&	\ion{C}{i}	&	11	&	-0.11	&	$\pm$0.09	&		&		&		\\													
N	&		&		&		&		&	SS	&	0.10	&	7.92	\\
	&	\ion{N}{i}	&	1	&	0.10	&		&		&		&		\\														
O	&		&		&		&		&	SS	&	-0.22	&	8.83	\\
	&	\ion{O}{i}	&	3	&	-0.22	&		&		&		&		\\															
Na	&		&		&		&		&	SS	&	0.03	&	6.33	\\
	&	\ion{Na}{i}	&	1	&	0.03	&		&		&		&		\\														
Mg	&		&		&		&		&	SS	&	-0.38	&	7.58	\\
	&	\ion{Mg}{i}	&	1	&	-0.38	&		&		&		&		\\														
Al	&		&		&		&		&	SS	&	0.23	&	6.47	\\
	&	\ion{Al}{i}	&	1	&	0.09	&		&		&		&		\\														
	&	\ion{Al}{i}	&	1	&	0.37	&		&		&		&		\\														
Si	&		&		&		&		&	SS	&	-0.24	&	7.55	\\
	&	\ion{Si}{i}	&	1	&	-0.21	&		&		&		&		\\														
 	&	\ion{Si}{i}	&	1	&	-0.25	&		&		&		&		\\														
  	&	\ion{Si}{i}	&	1	&	-0.25	&		&		&		&		\\														
S	&		&		&		&		&	SS	&	-0.06	&	7.33	\\
	&	\ion{S}{i}	&	4 	&	-0.06	&		&		&		&		\\													
K	&		&		&		&		&	EW	&	-0.17	&	5.12	\\
	&	\ion{K}{i}	&	1	&	-0.17	&		&		&		&		\\													
Ca	&		&		&		&		&	EW	&	-0.21	&	6.36	\\
	&	\ion{Ca}{i}	&	12	&	-0.21	&	$\pm$0.04	&		&		&		\\
	&	\ion{Ca}{ii}	&	2	&	-0.20	&	$\pm$0.26	&		&		&		\\													
Sc	&		&		&		&		&	SS	&	-0.42	&	3.17	\\
	&	\ion{Sc}{ii}	&	1	&	-0.42	&		&		&		&		\\													
Ti	&		&		&		&		&	EW	&	-0.24	&	5.02	\\
	&	\ion{Ti}{i}	&	3	&	-0.42	&	$\pm$0.07	&		&		&		\\
	&	\ion{Ti}{ii}	&	9	&	-0.18	&	$\pm$0.08	&		&		&		\\													
V	&		&		&		&		&	SS	&	0.03	&	4.00	\\
	&	\ion{V}{ii}	&	2	&	0.03	&		&		&		&		\\														
Cr	&		&		&		&		&	EW	&	-0.21	&	5.67	\\
	&	\ion{Cr}{i}	&	13	&	-0.25	&	$\pm$0.09	&		&		&		\\
	&	\ion{Cr}{ii}	&	15	&	-0.17	&	$\pm$0.07	&		&		&		\\													
Mn	&		&		&		&		&	SS	&	-0.23	&	5.39	\\
	&	\ion{Mn}{i}	&	2	&	-0.23	&		&		&		&		\\											
Fe	&		&		&		&		&	EW	&	-0.12	&	7.50	\\
	&	\ion{Fe}{i}	&	62	&	-0.12	&	$\pm$0.06	&		&		&		\\
	&	\ion{Fe}{ii}	&	15	&	-0.12	&	$\pm$0.08	&		&		&		\\												
Co	&		&		&		&		&	SS	&	0.11	&	4.92	\\
	&	\ion{Co}{i}	&	1	&	0.11	&		&		&		&		\\
 	&	            &	1	&	0.10	&		&		&		&		\\													
Ni	&		&		&		&		&	EW	&	0.26	&	6.25	\\
	&	\ion{Ni}{i}	&	31	&	0.25	&	$\pm$0.08	&		&		&		\\
	&	\ion{Ni}{ii}	&	3	&	0.41	&	$\pm$0.10	&		&		&		\\											
Cu	&		&		&		&		&	SS	&	0.23	&	4.21	\\
	&	\ion{Cu}{i}	&	1	&	0.23	&		&		&		&		\\
Zn	&		&		&		&		&	EW	&	0.56	&	4.60	\\
	&	\ion{Zn}{i}	&	4	&	0.56	&	$\pm$0.06    &		&		&	\\
Y	&		&		&		&		&	SS	&	0.59	&	2.24	\\
	&	\ion{Y}{ii}	&	1	&	0.47	&		&		&		&		\\
	&	\ion{Y}{ii}	&	1	&	0.70	&		&		&		&		\\																															
\hline
\end{tabular}
\end{table}

\begin{table}
	\centering
	\contcaption{}
	\label{tab:tab02b}
	\begin{tabular}{lccccccc} 
	\hline
\bf{Elm.}	& \bf{Ion}	&	\bf{n}	&	\bf{$\epsilon$}	& \bf{$\sigma$}	&	\bf{Mtd.}	&	\bf{$\epsilon_*$}	&	\bf{$\epsilon_{\sun}$}	\\
\hline	
Zr	&		&		&		&		&	SS	&	0.51	&	2.60	\\
	&	\ion{Zr}{ii}	&	1	&	0.51	&		&		&		&		\\																	
Ba	&		&		&		&		&	SS	&	1.09	&	2.13	\\
	&	\ion{Ba}{ii}	&	1	&	1.09	&		&		&		&		\\			
La	&		&		&		&		&	SS	&	0.57	&	1.17	\\
	&	\ion{La}{ii}	&	1	&	0.57	&		&		&		&		\\														
Ce	&		&		&		&		&	EW	&	0.55	&	1.58	\\
	&	\ion{Ce}{ii}	&	5	&	0.55	&	$\pm$0.12	&		&		&		\\														
Nd	&		&		&		&		&	SS	&	0.69	&	1.50	\\
	&	\ion{Nd}{ii}	&	1	&	0.62	&		&		&		&		\\
	&		&	1	&	0.75	&		&		&		&		\\														
Sm	&		&		&		&		&	SS	&	0.86	&	1.01	\\
	&	\ion{Sm}{ii}	&	1	&	0.86	&		&		&		&		\\														
Eu	&		&		&		&		&	SS	&	0.40	&	0.51	\\
	&	\ion{Eu}{ii}	&	1	&	0.40	&		&		&		&		\\
	\hline
	\end{tabular}
\end{table}

Table~\ref{tab:tab03} illustrates the sensitivity of the abundances to variations in the atmospheric parameters for the analyzed species. The range of changes in the atmospheric parameters is determined based on the variation in iron abundance resulting from these parameter changes. Each atmospheric parameter is systematically changed until the change in iron abundance surpasses the standard deviation of the accepted abundance value. Therefore, the iron ionization balance remains valid within the parameter range outlined in Table~\ref{tab:tab03}.

As depicted in the table, most elements exhibit minimal sensitivity to changes in model parameters, with their sensitivities even lower than the standard deviations of the abundances. The only exception is barium, for which abundance was derived from strong \ion{Ba}{ii} 5854 {\AA} line which is sensitive to microturbulent velocity, raising concerns about its accuracy. All other variations fall well below acceptable error margins.

\begin{table}
	\centering
	\caption{Sensitivities of the derived abundances to changes in the atmospheric parameters. The column indicated as `Total' represents the square root of the sum of the squares of all changes for the given species.}
	\label{tab:tab03}
	\begin{tabular}{lcccc} 
	\hline
\bf{Ion} & \bf{$\Delta$$\textit{T}$\textsubscript{{\sc eff}} (K)}  & \bf{$\Delta$log {\textit {g}}} &  \bf{$\Delta$$\xi$ (km s$^{-1}$)} & \bf{Total} \\ 
& (+100/-100) & (+0.2/-0.2) & (+0.7/-0.5) & (+/-) \\
        \hline
\ion{Li}{i} 	&	0.08	/	-0.08	&	-0.02	/	0.02	&	0.00	/	0.00	&	0.08	/	0.08	\\
\ion{C}{i}	&	0.02	/	-0.02	&	0.01	/	-0.01	&	-0.02	/	0.01	&	0.03	/	0.02	\\
\ion{N}{i}	&	-0.02	/	0.02	&	0.03	/	-0.03	&	-0.03	/	0.02	&	0.05	/	0.04	\\
\ion{O}{i}	&	-0.02	/	0.03	&	0.03	/	-0.03	&	-0.01	/	0.01	&	0.04	/	0.04	\\
\ion{Na} {i}&	0.06	/	-0.06	&	-0.03	/	0.03	&	-0.01	/	0.00	&	0.07	/	0.07	\\
\ion{Mg}{i} &	0.06	/	-0.06	&	-0.02	/	0.02	&	-0.01	/	0.01	&	0.06	/	0.06	\\
\ion{Al}{i} &	0.06	/	-0.06	&	-0.03	/	0.03	&	0.00	/	0.00	&	0.07	/	0.07	\\
\ion{Si}{i} &	0.05	/	-0.06	&	-0.02	/	0.02	&	0.00	/	0.00	&	0.05	/	0.06	\\
\ion{S}{i} &	0.05	/	-0.04	&	0.00	/	0.00	&	-0.01	/	0.01	&	0.05	/	0.04	\\
\ion{K}{i} &	0.08	/	-0.07	&	-0.02	/	0.04	&	0.05	/	-0.05	&	0.10	/	0.09	\\
\ion{Ca}{i} &	0.08	/	-0.07	&	-0.02	/	0.03	&	-0.06	/	0.07	&	0.10	/	0.10	\\
\ion{Ca}{ii} &	0.00	/	-0.01	&	0.03	/	-0.04	&	-0.04	/	0.03	&	0.05	/	0.05	\\
\ion{Sc}{ii} &	0.05	/	-0.04	&	0.06	/	-0.05	&	-0.09	/	0.10	&	0.12	/	0.12	\\
\ion{Ti}{i} &	0.08	/	-0.08	&	-0.02	/	0.02	&	-0.02	/	0.01	&	0.08	/	0.08	\\
\ion{Ti}{ii} &	0.04	/	-0.03	&	0.06	/	-0.05	&	-0.08	/	0.11	&	0.11	/	0.12	\\
\ion{V}{ii} &	0.04	/	-0.04	&	0.06	/	-0.05	&	-0.02	/	0.02	&	0.07	/	0.07	\\
\ion{Cr}{i} &	0.08	/	-0.06	&	-0.01	/	0.03	&	0.00	/	0.03	&	0.08	/	0.07	\\
\ion{Cr}{ii} &	0.02	/	-0.02	&	0.06	/	-0.06	&	-0.01	/	0.09	&	0.06	/	0.11	\\
\ion{Mn}{i} &	0.08	/	-0.07	&	-0.02	/	0.02	&	-0.04	/	0.01	&	0.09	/	0.07	\\
\ion{Fe}{i} &	0.07	/	-0.07	&	-0.02	/	0.02	&	-0.06	/	0.06	&	0.09	/	0.09	\\
\ion{Fe}{ii} &	0.02	/	-0.01	&	0.06	/	-0.05	&	-0.04	/	0.05	&	0.07	/	0.07	\\
\ion{Co}{i} &	0.07	/	-0.07	&	-0.02	/	0.02	&	0.00	/	0.00	&	0.07	/	0.07	\\
\ion{Ni}{i} &	0.07	/	-0.07	&	-0.02	/	0.02	&	-0.06	/	0.06	&	0.09	/	0.09	\\
\ion{Ni}{ii} &	0.01	/	-0.01	&	0.06	/	-0.06	&	-0.05	/	0.05	&	0.08	/	0.08	\\
\ion{Cu}{i} &	0.09	/	-0.09	&	-0.02	/	0.02	&	-0.01	/	0.01	&	0.09	/	0.09	\\
\ion{Zn}{i} &	0.07	/	-0.06	&	-0.01	/	0.02	&	-0.10	/	0.13	&	0.12	/	0.14	\\
\ion{Y}{ii}	&	0.05	/	-0.04	&	0.05	/	-0.05	&	-0.05	/	0.07	&	0.09	/	0.09	\\
\ion{Zr}{ii} &	0.04	/	-0.04	&	0.06	/	-0.05	&	-0.02	/	0.03	&	0.07	/	0.07	\\
\ion{Ba}{ii} &	0.10	/	-0.09	&	0.02	/	-0.02	&	-0.37	/	0.47	&	0.38	/	0.48	\\
\ion{La}{ii} &	0.08	/	-0.07	&	0.04	/	-0.03	&	0.00	/	0.00	&	0.09	/	0.08	\\
\ion{Ce}{ii} &	0.07	/	-0.07	&	0.04	/	-0.04	&	-0.03	/	0.03	&	0.09	/	0.09	\\
\ion{Nd}{ii} &	0.09	/	-0.08	&	0.04	/	-0.03	&	-0.01	/	0.01	&	0.10	/	0.09	\\
\ion{Sm}{ii} &	0.08	/	-0.08	&	0.04	/	-0.04	&	-0.01	/	0.01	&	0.09	/	0.09	\\
\ion{Eu}{ii} &	0.08	/	-0.07	&	0.04	/	-0.04	&	0.00	/	0.00	&	0.09	/	0.08	\\
    \hline
	\end{tabular}
\end{table}

\section{Discussion}

\subsection{Comparison of the Results to Previous Analyses}

\begin{figure}
	\includegraphics[width=\columnwidth]{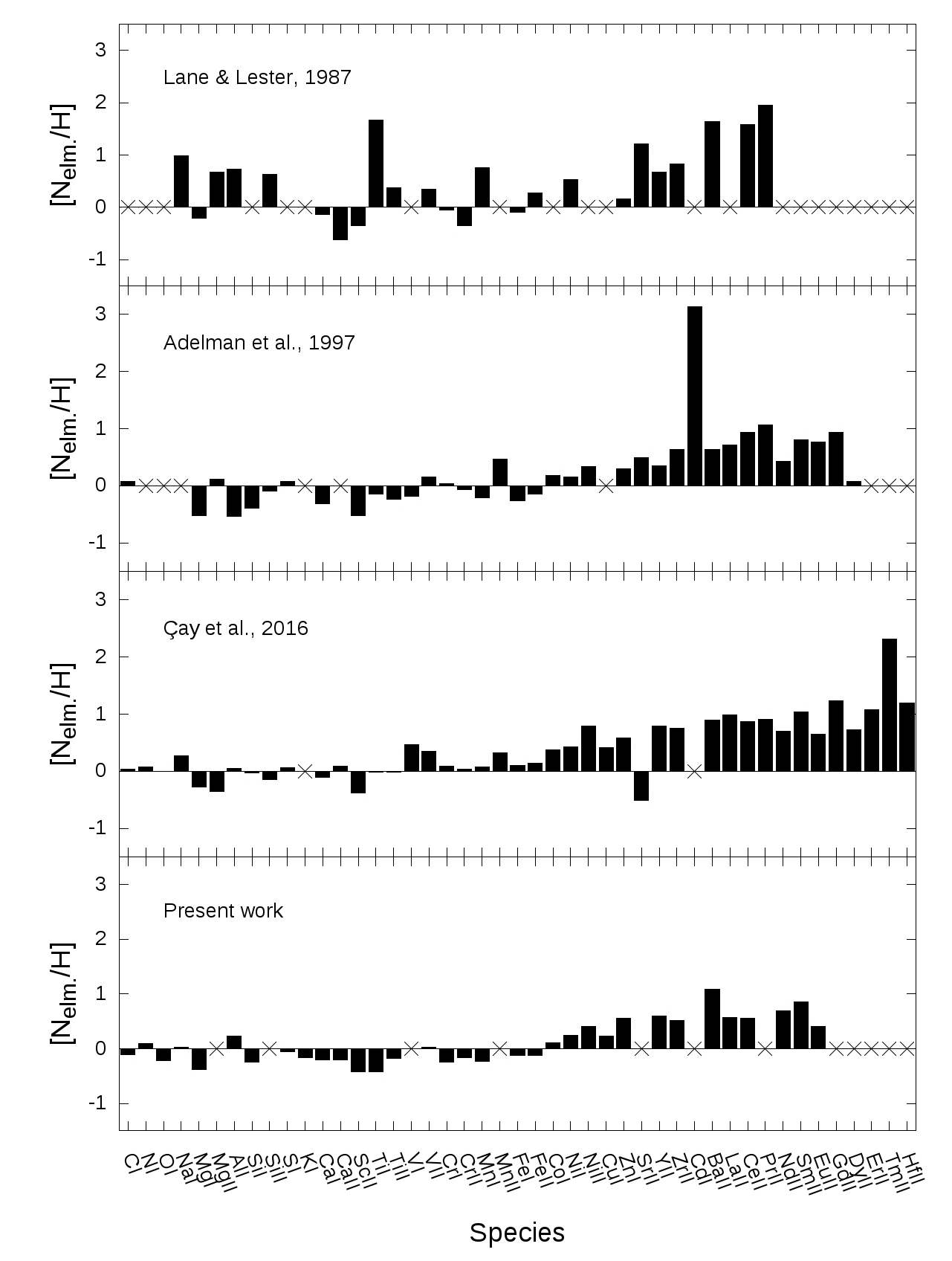}
    \caption{Comparison of the results of the present work to previous comprehensive analyses. Each work is compared for the common species with the present work. For comparison purposes, the abundance values of the previous works are not presented as the original ones but are recalculated relative to the solar values given by \citet{gre98}. Crosses indicates that the relevant species did not involve in the analyses.}
    \label{fig:fig06}
\end{figure}

Fig.~\ref{fig:fig06} compares our results to those of the previous comprehensive abundance analyses of the star, which use similar techniques and data. 

The inconsistency between our results and that of \citet{lan87} appears to stem from their use of very strong lines in the analysis, as well as differences in the model atmosphere and atomic parameters they used.

\begin{figure}
	\includegraphics[width=\columnwidth]{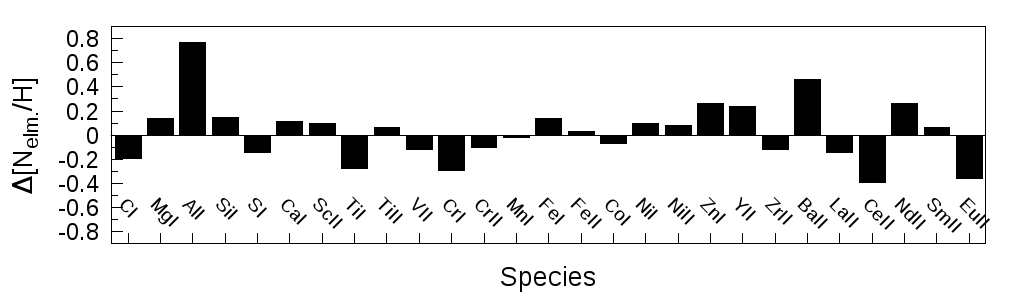}
    \caption{Differences in the abundances of the species common to \citet{ade97} (A97) and present work (PW) as $\Delta$$[N_{elm.}/H]$=$[N_{elm.}/H]$${_{PW}}$ - $[N_{elm.}/H]$${_{A97}}$.}
    \label{fig:fig07}
\end{figure}

However, distribution of abundances given by \citet{ade97} looks similar to ours. Fig.~\ref{fig:fig07} shows differences in the abundances of the 27 species common to their and our analyses. As seen from the figure, the differences are not systematic, which may be interpreted as they are resulting from the use of different atomic and model parameters. The selected lines and the number of lines involved can also be a cause of the differences. The figure shows that 52 per cent of the abundances match ours within a 0.1–0.2 dex difference, 15 per cent differ by less than 0.1 dex, while 33 per cent show significant differences.

\begin{figure}
	\includegraphics[width=\columnwidth]{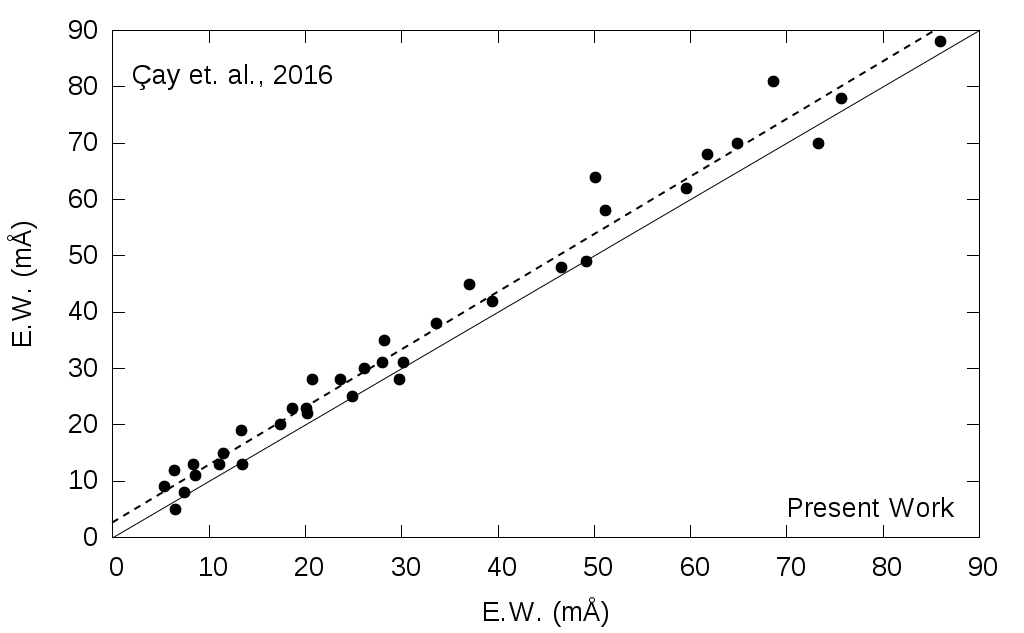}
    \caption{Comparison of our equivalent widths of \ion{Fe}{i} lines to that of common to \citet{cay16} clearly shows the systematic difference between two works. The dashed line is the least square fit. The difference is also getting larger toward the larger equivalent widths.}
    \label{fig:fig08}
\end{figure}

Fig.~\ref{fig:fig06} illustrates that the abundances determined in the more recent work by \citet{cay16} are generally higher than those reported by both \citet{ade97} and our findings. They also indicate that their results show an overall increase of 15 percent in abundances compared to the findings of \citeauthor{ade97}. We think that this general enhancement of all abundances may be due to some kind of systematic bias. Fig.~\ref{fig:fig08} compares their equivalent width measurements of common \ion{Fe}{i} lines to ours. The comparison shows that their equivalent widths are slightly larger than our measurements. Additionally, the number of lines they used in the analysis for any element is more than we used, which may have led to the inclusion of poorly resolved blended lines in the analysis. Indeed, our quick inspection of their line lists using our line selection procedure explained in section~\ref{sec:atmospheric} revealed the inclusion of severely blended lines in their analysis. It is not clear from the paper that what extent these lines were properly deblended without the help of spectrum synthesis.

Apart from these comprehensive abundance analyses of the star, \citet{burk91} specifically focused on the abundances of Li, Al, and Si, while \citet{tak12} concentrated on Li, Na, and K, and \citet{tak18} examined the abundances of C, N, and O. The following discussion considers the results of these authors on an element by element basis.

\subsubsection{Li}

Lithium abundance can be particularly important as it may indicate the evolutionary status of a star \citep[see, e.g.][]{burk91}. From measurements of the Li 6707 {\AA} line, \citeauthor{burk91} found the Li abundance of the star to be $\log{(\rm Li/H)}$ = 3.20, consistent with cosmic lithium abundance, $\log{\rm (Li/H)}$ = 3.10. The more recent work of \citet{tak12} found $\log{\rm (Li/H)}$ = 3.03 from the same line after applying their -0.08 dex NLTE correction; thus, their LTE value must be $\log{\rm (Li/H)}$ = 3.11, which is in good agreement with \citeauthor{burk91}. We also used the Li 6707 {\AA} line in our analysis and found $\log{\rm (Li/H)}$ = 3.20, confirming the results of both analyses.

\subsubsection{C,N,O}
\label{sec:CNO}

Carbon, nitrogen and oxygen abundances are another indicator of evolutionary status of a star. \citet{tak18} reports NLTE abundances for carbon, nitrogen, and oxygen relative to Procyon, with [C/H] = -0.17, [N/H] = -0.22, and [O/H] = -0.18. These values correspond to [C/H] = 0.05, [N/H] = 0.18, [O/H] = -0.14 relative to solar values given by \citet{gre98}, after NLTE corrections are subtracted from the results to compare with our LTE results. Their C, N, and O abundances are 0.16, 0.08, and 0.08 dex higher than ours, respectively. Those differences may well be explained for nitrogen and oxygen by the use of different atomic and model parameters. However, their carbon abundance is slightly larger than our result and closer to those of \citet{ade97}. Nevertheless, both studies used only the \ion{C}{i} 5380 {\AA} line in their analyses, which is also included in our analysis and consistently gives a higher abundance in all three works, while the other lines we used yield underabundant values in our analysis.

\subsubsection{Na}

Our result from the \ion{Na}{i} 6154 {\AA} line is in good agreement with \citet{tak12}, which gives $\log{(Na/H)}$ = 6.31. \citeauthor{tak12} used the \ion{Na}{i} 5682 {\AA} and 5688 Å lines for their calculations, which have corresponding NLTE corrections of -0.07 and -0.08 dex, respectively. Their LTE [Na/H] value differs by only 0.02 - 0.03 dex from our sodium abundance estimation.

\subsubsection{Al}

\citet{burk91} suggest using the 6696  {\AA} and 6698 {\AA} lines of neutral aluminum instead of the strong 3944  {\AA} and 3961  {\AA} lines, referencing the work of \citet{bur89} on Hyades cluster stars. The aluminum abundance of 15 Vul, as reported by \citeauthor{burk91}, corresponds to [Al/H] = 0.23 when \citet{gre98} solar values are considered. In contrast, \citet{ade97} used the blue lines in their equivalent width analysis, with their abundance value corresponding to [Al/H] = -0.58 on average.

In Table~\ref{tab:tab02a}, we have presented the results obtained from the 6696 {\AA} and 6698 {\AA} lines. We also calculated the aluminum abundance by synthesizing the 3961 {\AA} line and obtained [Al/H] = -0.56, showing a striking difference between the calculation with the blue line and the red lines. Our calculations confirm the results of both \citeauthor{burk91} and \citeauthor{ade97}. However, we favor the suggestion of \citeauthor{burk91} because the red lines are not too strong for abundance derivation. On the other hand, the 3961 {\AA} line is too strong and very sensitive to changes in microturbulent velocity (its sensitivity to $\Delta$$\xi$ is -0.38/+0.31, while sensitivity to $\Delta$\textit{T}\textsubscript{{\sc eff}} is +0.08/-0.08, and to $\Delta$log {\textit {g}} is +0.01/-0.01, to compare to the sensitivity of the red lines given in Table~\ref{tab:tab03}). This issue raises concerns about the accuracy of the abundance obtained from this line. 

\subsubsection{Si}

\citet{ade97} used very strong and blended lines of \ion{Si}{i} and \ion{Si}{ii} to derive the silicon abundance, with their conclusions for each stages being [Si/H] = -0.43 and [Si/H] = -0.14, respectively, relative to the solar values given by \cite{gre98}. However, the blending and strength of their lines raise some questions about their results. \citet{burk91} used the \ion{Si}{i} 6722 {\AA} line and obtained [Si/H] = -0.15 (although it is given as -0.1 in their table 4).

We have been able to use different and cleaner lines compared to these authors, taking advantage of lines from the red part of the spectrum. The lines listed in Table~\ref{tab:tab01} provide perfectly consistent results, averaging around [Si/H] = -0.24. We did not include the 6722 {\AA} line used by \citeauthor{burk91} in our analysis list because the line is closer to the edge of the \'{e}chelle order in our spectrum. Nevertheless, when we synthesized this line for cross-checking, we found that it also matches the synthetic line produced with our mean value of -0.24. However, \citeauthor{burk91}'s log {\textit {gf}} value of -1.19, as given in \citet{bur89}, does not properly reproduce the solar \ion{Si}{i} 6722 {\AA} line when using \citeauthor{gre98} as the abundance reference. Because of that, we used a log {\textit {gf}} value of -1.07 obtained from inverse analysis of the line while performing our cross-check.

The only \ion{Si}{ii} line suitable to our selection criteria was the 6371 {\AA} line. Nevertheless, this line gives a value of [Si/H] = 0.28, which is 0.52 dex larger than the calculated from the \ion{Si}{i} lines. Sensitivity analysis of the model parameters showed that the line is slightly sensitive to microturbulence. It may also have issues related to photoionization cross-sections and NLTE effects \citep[see][]{mas22}, which could lead to an overestimated abundance. To stay on the safe side, we excluded this line from our analysis.

\subsubsection{K}

Potassium is one of the rarely analyzed elements in A-type stars because it requires analyzing very weak lines of the element, if detectable \citep{tak12}.

\begin{figure}
	\includegraphics[width=\columnwidth]{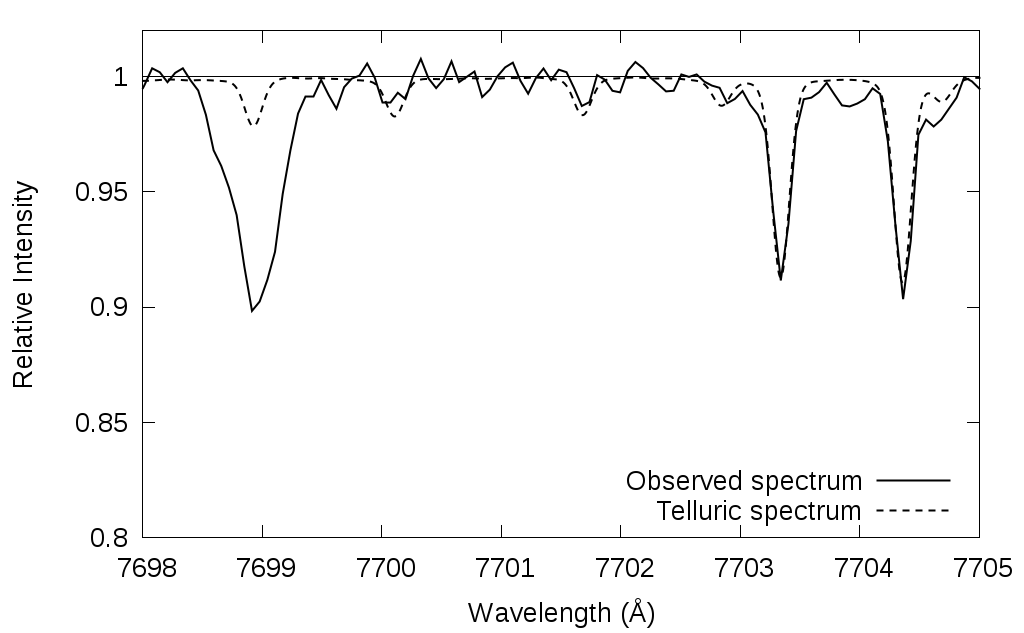}
    \caption{The \ion{K}{i} 7699 {\AA} line is used to determine the potassium abundance. This line is blended with a telluric line. Two nearby telluric lines, around 7704 {\AA}, are shown in the figure to demonstrate how well the telluric spectrum (dashed line), extracted from the Solar Flux Atlas of \citet{wal11}, matches the telluric lines in the stellar spectrum (continuous line). See the text for details of the calculation.}
    \label{fig:fig09}
\end{figure}

\ion{K}{i} 7699 {\AA} line was the only detectable line in our spectrum, which allowed us to derive the potassium abundance from it. As seen in Fig.~\ref{fig:fig09}, the line contains a considerable telluric component. A typical and correct approach for analyzing this line would involve removing the telluric contribution by dividing the observed spectrum by the telluric spectrum. However, because the telluric spectrum we used was not part of our data, applying this procedure was not possible. Instead, we measured the equivalent widths of the telluric line and the potassium line separately and subtracted the equivalent width of the telluric line from that of the potassium line as our best attempt. This procedure might be plausible since the telluric lines match our spectrum well (see Fig.~\ref{fig:fig09}). The derived relative potassium abundance is [K/H] = -0.17. This value closely matches the [K/H] = -0.24 LTE result, after removing the -0.28 dex NLTE correction and applying the conversion, as explained previously, given by \citeauthor{tak12}, who used the same line for analysis. The difference between the two results is quite acceptable, at only 0.07 dex.

\subsection{Evaluation of the Atmospheric Parameters}
\label{sec:Evaluation of the Atmospheric Parameters}
Table~\ref{tab:tab04} summarizes the atmospheric parameter estimations of 15 Vul found in the literature, including those from the present work. Since the iron ionization balance remains valid and the variations in abundances are within their standard deviations across the range of changes in the model parameters listed in Table~\ref{tab:tab03}, this range could also be considered the acceptable error range for these parameters. Therefore, from Table~\ref{tab:tab04}, our effective temperature estimation remains within the estimations of the previous works. However, this is not the case for surface gravity and microturbulent velocity.

In the table, the lowest surface gravity and microturbulent velocity values are from our study. As explained in section~\ref{sec:parameters}, we calculated the initial atmospheric parameters using Str\"{o}mgren photometry and iteratively tuned the parameters by adjusting the values to achieve excitation and ionization equilibria of iron. As seen in the Table~\ref{tab:tab04}, most works accept the photometry- and spectrophotometry-based parameters without further tuning, which seems to be the main cause of the differences. In this context, one might ask why the results of other works that used a similar methodology of fine-tuning differ.

\begin{table*}
	\centering
	\caption{Atmospheric parameter estimates of 15 Vul present in the literature and found in this study. The letters in the brackes indicates the estimation methods; C: curves of growth, E: spectral energy distribution, A: standart abundance analysis, P: photometry, S: spectrum synthesis, B: Blackwell diagrams.}
	\label{tab:tab04}
	\begin{tabular}{lccl} 
	\hline
\bf{$\textit{T}$\textsubscript{{\sc eff}}}	& \bf{log {\textit {g}}} & \bf{$\xi$}  &	\bf{Ref.}	\\
(K) & & (km s$^{-1}$) & \\
\hline
8000 & 3.50 (C) & 4.50 (C) & \citet{far71} \\
7500 & 3.50 (E) & 3.40 (A) & \citet{lan87} \\
8040 & 3.70 (P) & 4.00 (A) & \citet{burk91} \\
7600 & 3.20 (S/A) & 4.00 (A) & \citet{bol92} \\
7700 & 3.50 (S/A) & 4.00 (A) & \citet{ade97} \\
7850 & 3.70 (P) & 4.10 (S/B) & \citet{lan09} \\
7870 & 3.62 (P) & 5.16 (S) & \citet{tak12} \\
7825 & 3.45 (S/A) & 2.80 (S/A) & \citet{cay16} \\
7685 & 3.09 (A) & 1.95(A/B) & Present work \\
    \hline
	\end{tabular}
\end{table*}

\begin{figure}
	\includegraphics[width=\columnwidth]{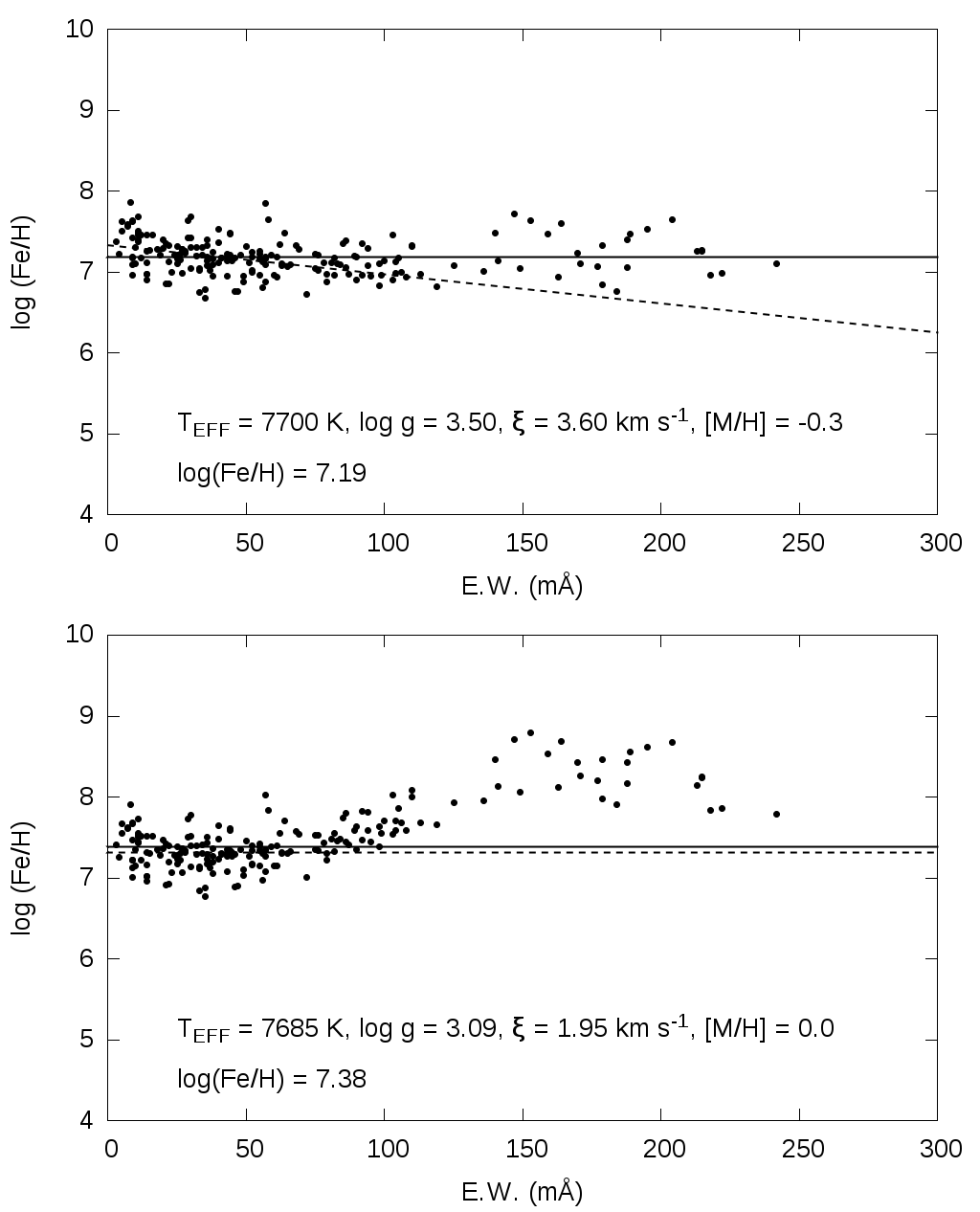}
    \caption{The figure in the upper panel is reproduced from \citet{ade97}'s analysis for \ion{Fe}{i} lines, using their equivalent width measurements, atomic values and microturbulance value from their table 1. $\And$ 2. , and an {\sc atlas9} model calculated using their model parameters. The continuous line represents the mean abundance they found from \ion{Fe}{i} lines, while the dashed line is the linear least-squares fit through the lines with an equivalent width of less than 100 m{\AA}. It appears that the authors obtained a mean abundance by considering very strong lines, which resulted in a trend among the weaker lines. The lower panel is similar, but our model is used, and only weak to moderately strong lines are considered for analysis. There is no trend in the weaker lines. Their equivalent width measurements and line data perfectly reproduce the abundance we derived from our data.}
    \label{fig:fig10}
\end{figure}

In the Table~\ref{tab:tab04}, the most suitable work for cross-checking is \citet{ade97}, because \citet{bol92} used {\sc atlas8} for their calculations and \citet{cay16} may have systematics as explained previously. Fig.~\ref{fig:fig10}, produced using \citeauthor{ade97}'s data, clearly shows the cause of the difference between the microturbulent velocities in the two works. From the upper panel of the figure, it appears that their use of strong lines in the analysis, and forcing them to agree with the abundance obtained from the weak lines, resulted in a higher microturbulent velocity value due to the Van der Waals broadeining affecting the strong lines (see, e.g., \citet{sma14} and references therein). Using the data of \citeauthor{ade97} but considering only the lines with equivalent widths lower than 100 m{\AA} results in a perfect fit to the model parameters and the iron abundance we obtained in the present work (see bottom panel of Fig.~\ref{fig:fig10}).

The highest microturbulent velocity in the Table~\ref{tab:tab04}, reported by \citet{tak12}, may be attributed to their specific procedure for determining microturbulent velocity and their use of the NLTE approach.

Our low surface gravity value requires some consideration. If the star exhibits $A_{m}$ star characteristics, the surface gravity which is obtained from Str\"{o}mgren photometry may be overestimated due to line blanketing affecting the $c_{1}$ passband \citep{cat12}. 

\begin{figure}
	\includegraphics[width=\columnwidth]{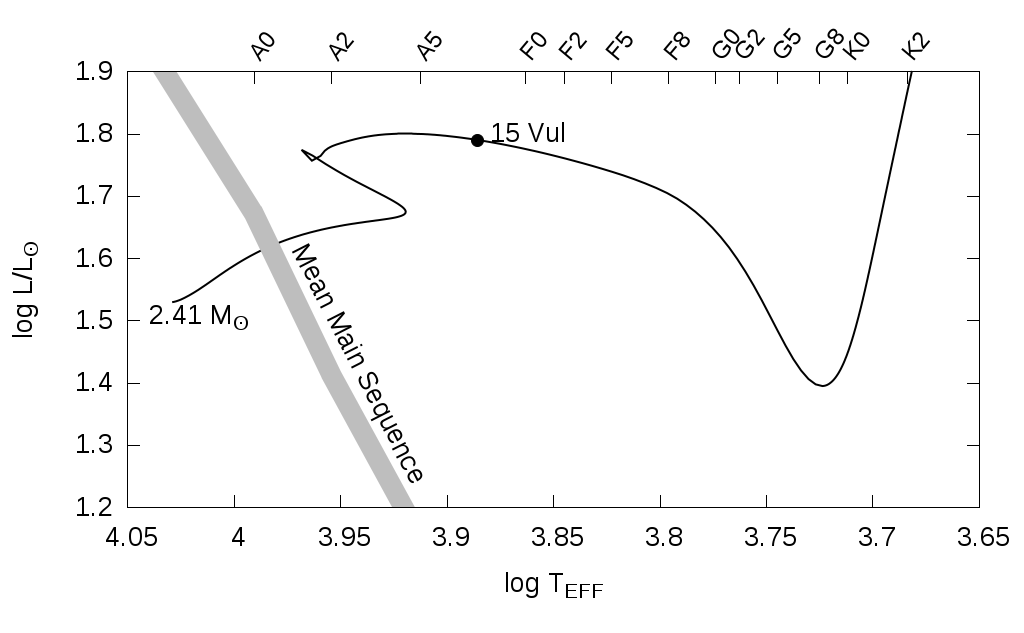}
    \caption{The star’s effective temperature and luminosity, with $\log L/L_{\sun}$ = 1.79, suggest a mass of 2.41 $M_{\sun}$ according to the BaSTI evolutionary track (solar scaled, no overshooting, no mass loss model with Z = 0.0172, Y = 0.2695). The star has apparently left the main sequence and is in the subgiant phase, around A7 spectral type. The location of the mean main sequence and spectral types is based on \citet{dri2000}.}
    \label{fig:fig11}
\end{figure}

It was noted by \citet{mic56} that the spectroscopic gravity of 8 Com, 15 Vul, and $\tau$ UMa is about ten times smaller than the gravity value based on luminosity. 
If we assume the star follows the evolutionary track of a solar-composition normal star, with no factors affecting its luminosity, an independent surface gravity determination can be obtained from Eq.~\ref{eq:surface gravitiy}. Using the parallax of 13.443 mas from the GAIA DR3 archive \footnote{\url {https://gea.esac.esa.int/archive/}}, calculating the luminosity from Eq.~\ref{eq:bolometric magnitude} and \ref{eq:luminosity} and estimating the star's mass from BaSTI \footnote{\url {http://basti-iac.oa-abruzzo.inaf.it/tracks.html}} tracks \citep{hid18} (see Fig.~\ref{fig:fig11}), we find log {\textit {g}} = 3.53, which is consistent with the value obtained from Str\"{o}mgren photometry. 

\begin{equation}
    \log {\textit g}_{\star}=\log {\textit g}_{\sun}+\log {\frac {M_{\star}}{M_{\sun}}}+4 \log {\frac {T_{{\rm eff}{\star}}}{T_{{\rm eff}{\sun}}}}+0.4 (M_{{bol.}{\star}}-M_{{bol.}{\sun}})
    \label{eq:surface gravitiy}
 \end{equation}

 \begin{equation}
    M_{bol.}=m_{v}+5-5 \times log \ {\frac{1}{\upi}}-A_{V} + B.C.
	\label{eq:bolometric magnitude}
 \end{equation}

 \begin{equation}
   M_{{bol.}{\star}}-M_{{bol.}{\sun}}=-2.5 \log{\frac {L_{\star}}{L_{\sun}}}
	\label{eq:luminosity}
 \end{equation}

Regardless of whether the star is considered an  $A_{m}$ star or a normal star, the discrepancy could also stem from using the iron ionization balance to determine surface gravity. For instance, the systematic difference between spectroscopic and photometric surface gravity values is a well-established issue for cooler stars \citep[see][]{mor14}.

However, using a log {\textit {g}} value of 3.09 instead of 3.53 would primarily affect the abundances derived from singly ionized lines. Nevertheless, according to the sensitivity of our lines to log {\textit {g}}, as presented in Table~\ref{tab:tab03}, this would result in changes to the abundances of less than 0.1 dex.

\subsection{The Abundance Pattern}
\begin{figure}
	\includegraphics[width=\columnwidth]{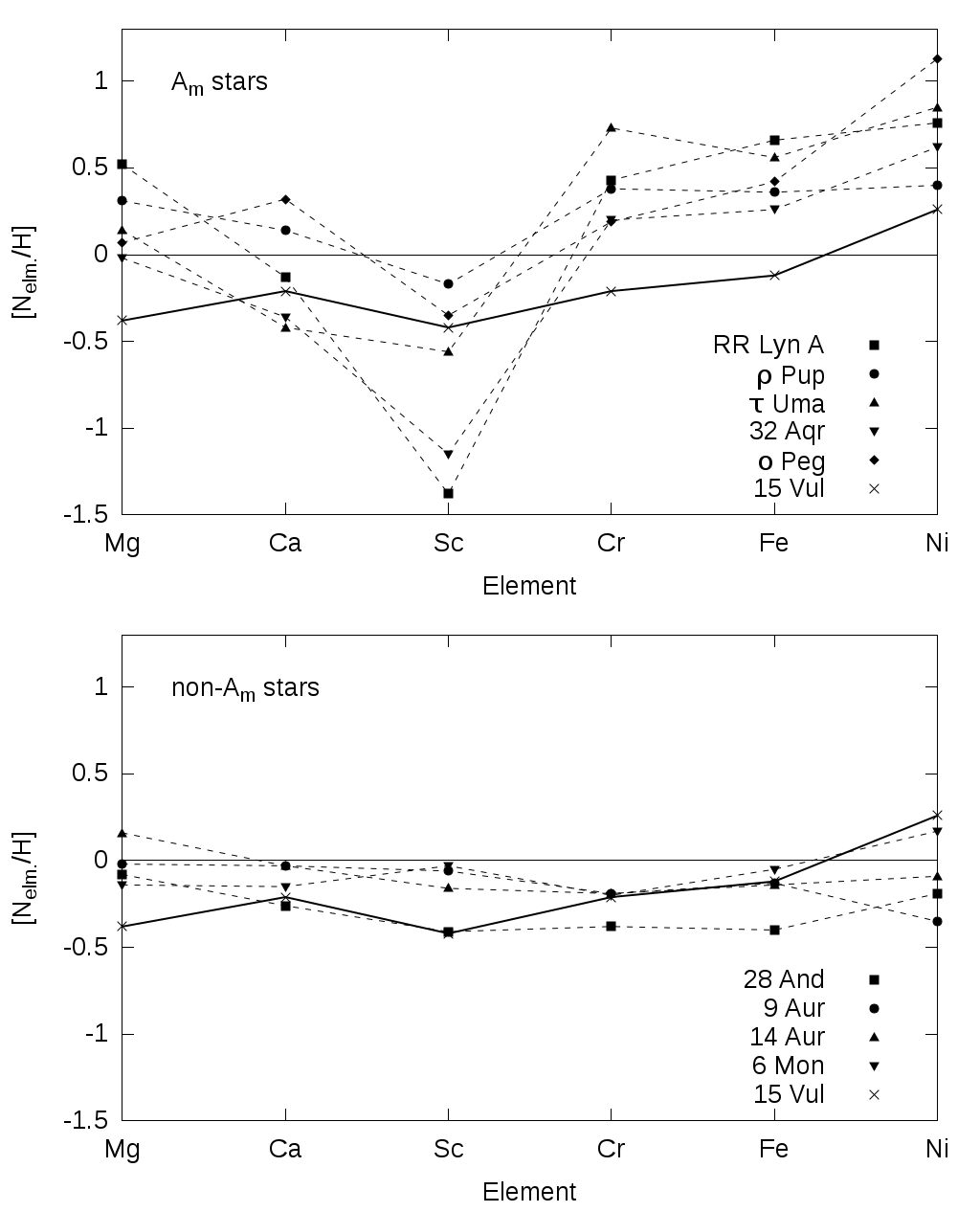}
    \caption{Comparison of the abundance pattern of 15 Vul (continuous line) with $A{_m}$ and non-$A{_m}$ stars (dashed lines) from \citet{hui00}. The abundance pattern of 15 Vul aligns with that of non-$A{_m}$ stars.}
    \label{fig:fig12}
\end{figure}

\begin{figure}
	\includegraphics[width=\columnwidth]{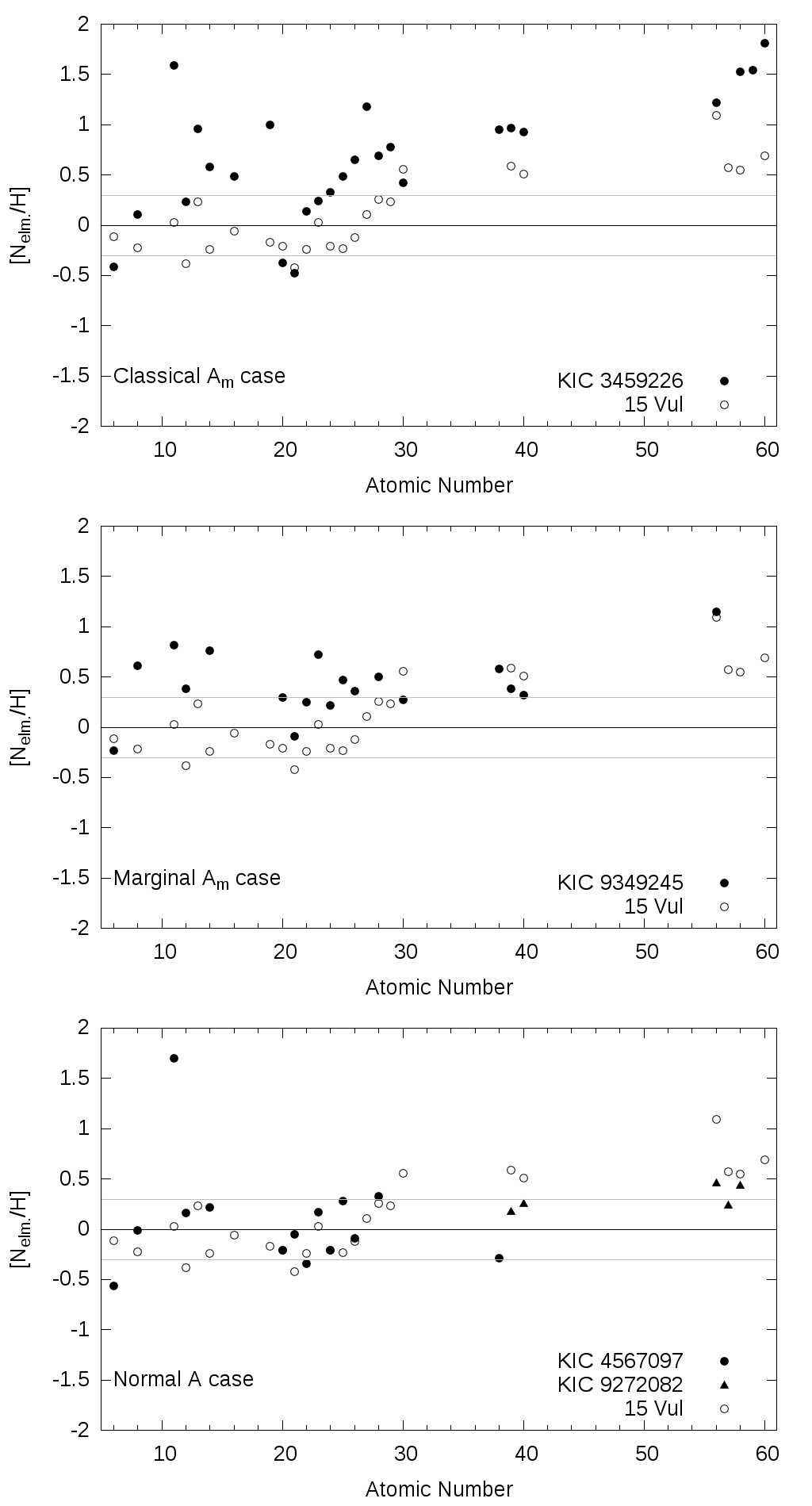}
    \caption{Comparison of the abundance pattern of 15 Vul with selected stars from \citet{tru21}. The abundance pattern of 15 Vul aligns with that of normal A stars. Because KIC 4567097 is lack of heavy element measuremets due to high rotation, heavy element abundances is completed from KIC 9272082 measurements. All abundances are recalculated relative to solar values from \citet{gre98}. Grey lines show (quasi) solar abundance range $\pm$0.3 dex from $[N_{elm.}/H]$ = 0 as adopted from \citet{hui00}.}
    \label{fig:fig13}
\end{figure}

Comparing the star's abundance pattern found in the present work to that of other A-type stars can be illustrative. \citet{hui00} presented abundances of Mg, Ca, Sc, Cr, Fe, and Ni for nine field A and $A{_m}$ stars and classified them as $A{_m}$ if the stars exhibited calcium and/or scandium deficiency and/or an overabundance of at least two iron peak elements (represented by Cr, Fe, and Ni in the study). If the calcium and scandium abundances were normal and two of the iron peak elements also showed normal abundances, the star was classified as a normal A-type star. An element was considered to have a (quasi) solar abundance if its abundance was within $\pm$0.3 dex of the solar value. Fig.~\ref{fig:fig12} is created from the data given by \citeauthor{hui00} with addition of our results for 15 Vul for comparison purposes. The abundance pattern we derived for 15 Vul closely matches that of non-$A{_m}$ stars. However, this similarity does not strongly indicate whether it is a marginal $A{_m}$ star or a normal A star. Fig.~\ref{fig:fig13}, is created in the same way from the data taken from \citet{tru21} for selected stars, and allows for a comparison across a broader set of elements. From this figure, the closest match is with the normal A star composition. Except for scandium, marginal $A{_m}$ stars show overabundances of iron peak elements, similar to classical $A{_m}$ stars but to a lesser extent. In contrast, normal A stars show no such overabundances for most of these elements, with iron itself being underabundant. The underabundance of iron in 15 Vul appears consistent across all studies we compared with our own findings, except for \citet{cay16}.

\section{Conclusions}


We conducted a detailed abundance analysis of 15 Vul, known as a marginal $A{_m}$ star, under the LTE assumption. This analysis employed a combination of equivalent width analysis and spectrum synthesis methods, using the most recent and carefully selected atomic parameters. We compared our results with previous analyses of the star, evaluating both their findings and ours, and provided improved abundance values. 

Additionally, we compared our results with those of other normal A-type and metallic-line stars. Based on our findings, we draw conclusions about the peculiarity and evolutionary status of the star as follows.

There is no uncertainty regarding our estimation of $\textit{T}$\textsubscript{{\sc eff}} = 7685 K, as it aligns well with the $\textit{T}$\textsubscript{{\sc eff}} values reported in the relevant literature. The different methods of $\textit{T}$\textsubscript{{\sc eff}} determination we employed also provide consistent results.

We employed two independent methods to determine the microturbulence velocity in the atmosphere, both consistently yielding a value of approximately 2 km s$^{-1}$. 

As discussed in subsection~\ref{sec:Evaluation of the Atmospheric Parameters}, our surface gravity value appears to be low. The log {\textit g} = 3.53 determined via parallax should be closer to the actual value. However, we chose to use log {\textit g} = 3.09 throughout the abundance analysis to maintain internal consistency within the model atmosphere. 

Our comparison of the abundance pattern of 15 Vul to other A-type stars leads us to conclude that it is a normal A star rather than a marginal $A{_m}$ star regarding to abundance pattern, except calcium and scandium. Our microturbulence estimation of approximately 2 km s$^{-1}$ further supports its normal star characteristics, as $A{_m}$ stars typically have higher microturbulence values around 4 km s$^{-1}$, whereas normal A stars have values closer to 2 km s$^{-1}$ \citep[see][]{lan09}.

On the other hand, the RUWE (Renormalized Unit Weight Error) value of 1.766 from the GAIA archive  suggests that the star has a very low-mass companion \citep{gai22}. Furthermore, the projected rotational velocity of 10 km s$^{-1}$ \citep{abt95} is unusually low for a normal A-type star. Both binarity and low rotational velocity are distinct characteristics of $A{_m}$ stars. However, it could also be argued that the low projected rotational velocity may be due to the star being observed nearly pole-on.

Regarding surface gravity, the star is in the subgiant phase, as indicated by the evolutionary tracks (see Fig.~\ref{fig:fig11}). This is also consistent with the CNO abundances of the star. \citet{ric00}'s predictions for CNO abundances in 2.5 $M_{\sun}$ stars (which closely matches the estimated mass of 15 Vul) suggest that carbon, nitrogen and oxygen are underabundant in $A{_m}$ stars but shift to solar levels during the subgiant phase due to the first dredge-up \citep[see][figure 15]{ric00}. Our CNO estimates show that the star has near-solar values, leading us to conclude that some convection process is active in 15 Vul. Even when NLTE corrections for carbon, nitrogen and oxygen are applied, as discussed in subsection~\ref{sec:CNO}, the abundances remain close to solar values. In this context, one might expect a depleted lithium abundance which is not the case for 15 Vul. However, the lithium abundance does not provide clear evidence about the star's evolutionary status, as all lithium deficient A stars are evolved, but not all evolved A stars show lithium deficiency \citep[see e.g.][and references therein]{nor05}.

Based on a statistical analysis of normal and chemically peculiar A-type stars, \citet{abt17} reaches a preliminary conclusion that `\textit{$A{_m}$ stars plus A4-F2 normal stars evolve into A7-F9 IV stars with normal abundances and then into F2-F9 (or later) III with normal abundances}'. The location of 15 Vul on the HR diagram aligns precisely with this prediction (see Fig.~\ref{fig:fig11}).

The final justification could be made by considering the abundances of calcium and scandium. These elements are underabundant in the star’s atmosphere compared to normal A stars, as seen in $A{_m}$ stars. \citet{leb08} deduced from their theoretical calculations that depending on the scenario that could occur in $A{_m}$ star, surface underabundances of Ca and Sc might be created in the shallow or deeper regions such that either of them would lead to Ca and Sc being in noble gas configurations, causing their underabundances. This situation will persist in the subgiant phase. The extent of overabundances of other elements will depend on the depth of the mixing zone when the star reaches the subgiant phase. 

Therefore, we may conclude from all these considerations that 15 Vul was a classical $A{_m}$ star on the main sequence and has evolved into the subgiant stage, where its abundance pattern has gradually changed to that of a (quasi) normal A star, retaining slow rotation. We further predict that this progression toward a normal abundance pattern will continue as the star evolves to around the late type giants where its rotational characteristic may no longer be distinguishable. Our results may provide the first observational evidence of \citeauthor{abt17}’s prediction. This conclusion could be coincidental, however, and should ideally be supported or tested with a large sample of stars with homogeneously obtained abundances.

\section*{Acknowledgements}

We thank T\"{U}B\.{I}TAK National Observatory for a partial support in using RTT150 (Russian-Turkish 1.5-m telescope in Antalya) with project number 11ARTT150-119.

This work has made use of;

- the SIMBAD database, CDS, Strasbourg Astronomical Observatory, France,

- the VALD3 database, operated at Uppsala University, the Institute of Astronomy RAS in Moscow, and the University of Vienna,

- the NIST Atomic Spectra Database,

- the Astrophysics Data System, funded by NASA under Cooperative Agreement 80NSSC21M00561.

- the BaSTI web tools,

- data from the European Space Agency (ESA) mission
{\it Gaia} (\url{https://www.cosmos.esa.int/gaia}), processed by the {\it Gaia} Data Processing and Analysis Consortium (DPAC, \url{https://www.cosmos.esa.int/web/gaia/dpac/consortium}). Funding for the DPAC has been provided by national institutions, in particular the institutions participating in the {\it Gaia} Multilateral Agreement.

The authors are grateful to T. Sitnova and I. F. Bikmaev for their valuable discussions, and to the anonymous referee for their constructive feedback and comments.

\section*{Data Availability}

The observational data will be shared on request to the corresponding author with permission of T\"{U}B\.{I}TAK National Observatory (TUG).

The full version of Table~\ref{tab:tab01} is also available at VizieR facility. URL: \url{https://vizier.cds.unistra.fr/viz-bin/VizieR}



\bibliographystyle{mnras}
\bibliography{15Vul} 








\bsp	
\label{lastpage}
\end{document}